\documentclass[10pt]{article}
\usepackage{graphicx}
\usepackage{amsmath}
\usepackage{lineno}
\usepackage{hyperref}
\makeatletter

\begin{document}
\title{\textbf{ Oscillatory Growth: A Phenomenological View}}
\author{\textbf{Dibyendu Biswas$^1$}\footnote{dbbesu@gmail.com} \textbf{, Swarup Poria$^2$}\footnote{swarupporia@gmail.com} \textbf{and Sankar Nayaran Patra$^3$}\footnote{sankar.journal@gmail.com}\\
\small{$^1$Department of Basic Science, Humanities and Social Science (Physics)}\\\small{ Calcutta Institute of Engineering and Management }\\
\small {Kolkata-700040, India}\\
\small{$^2$Department of Applied Mathematics, University of Calcutta}\\
\small {Kolkata-700009, India}\\
\small{$^3$ Department of Instrumentation Science, Jadavpur University}\\
\small {Kolkata-700032, India}}

\date{}
\maketitle

{\bf{Abstract}}\\
In this communication, the approach of phenomenological universalities of growth are considered to describe the behaviour of a system showing oscillatory growth. Two phenomenological classes are proposed to consider the behaviour of a system in which oscillation of a property may be observed. One of them is showing oscillatory nature with constant amplitude and the other represents oscillatory nature with a change in amplitude. The term responsible for damping in the proposed class is also been identified. The variations in the nature of oscillation with dependent parameters are studied in detail. In this connection, the variation of a specific growth rate is also been considered. The significance of presence and absence of each term involved in phenomenological description are also taken into consideration in the present communication. These proposed classes might be useful for the experimentalists to extract characteristic features from the dataset and to develop a suitable model consistent with their data set.\\

\textbf{Keyword:}\\

Phenomenological universality, Oscillation in growth, Damping, Specific growth rate.
\section{Introduction}
In the field of science and technology, it is found that many characteristic patterns have been discovered that are remarkably similar in nature, though they are related to totally different phenomenologies {\cite {Carpinteri2002,Carpinteri2005,Carpinteri1994,Guyer,Delsanto2007}}. It is because of the fact that the background mathematical formalism remains the same. Therefore, an universality in terms of mathematical formalism is observed, in which a mathematical equation is showing universality in the sense that the variables involved in the equation may indicate a scientific discipline. The opposite may also be true in the sense that some related effect might be found in nature for every mathematical niche, that is generally observed in biology. However, this might be found in the other field of application. For example, any symmetry class, predicted mathematically, is found to be available in nature. Therefore, a search for universality in different phenomenologies is very important from the point of view of applications. A cross-fertilization among different fields might be accelerated when different fields share a common mathematical formalism or concept. One of such approaches is the classification scheme of phenomenological universalities {\cite {Castorina}}.\\
In the phenomenological approach, the universality class at the level $n=0$ corresponds to Malthusian or autocatalytic processes. At the level $n=1$, it produces Gompertz law of growth {\cite {Gompertz}} that was being used for more than a century to describe growth phenomena in diversified fields. A variation of the class, termed as involuted Gompertz function, is used to describe the evolution and involution of organs undergoing atrophy {\cite {Molski}}. The class corresponding to the level $n=2$ leads to the growth equation, proposed by West $et.$ $al.$ to describe different biological growth patterns {\cite {West1997,West1999,Brown,West2004,West2001}}. The classes corresponding to $n=1$ and $n=2$ have been applied in diversified and unrelated fields {\cite {Delsanto2008,Delsanto2009,Pugno,Gliozzi}}. An extension of this approach, termed as complex universality, was characterized {\cite {Delsanto2011}} and applied to explain the concurrent growth of two phenotypic features {\cite {Barberis2010}}. A vector generalization of the phenomenological universality is used to describe the interactive growth of two or more organisms in a given environment {\cite {Barberis2011,Barberis2012}}.\\
The formalism of phenomenological approach {\cite {Castorina}} of classes n= 0, 1, 2 leads to monotonous growth of the system with no scope of oscillations, though they are an essential feature of different complex systems of nature. In some cases they are not observed, might be due to the fact that they were overlooked, or masked by large experimental errors or missed due to the inadequate range of variation of the variables being measured. Apart from these factors, oscillations are observed in different types of complex systems {\cite {Friedman,Huisman,Sibona,Kruse,Latif,pullela,murray,kroeger,tu2007,dane,tu2005}}. The existence of volume oscillation, of relatively large amplitude, in the growth of multicellular tumour spheroids was experimentally found {\cite {Deisboeck,Chignola}}. A variation of the phenomenological approach in the complex field, instead of real field, termed as complex universalities is applied to analyze two correlated features of the growing oscillatory systems {\cite {Delsanto2011}}. In the same framework, concurrent growth of phenotypic features are also explained and the correlation between phenotypic features is derived by Barberis $et.$ $al.$ {\cite {Barberis2010}}.One of essential features of the complex universalities approach is that the growth or decay in the amplitude is unavoidable{\cite {Delsanto2011}}. Therefore, it is not possible to explain the oscillating behaviour of a system with constant amplitude.\\
In the phenomenological approach, it is expected that the values of the coefficients are governed by the particular growth mechanism involved in the system {\cite{Castorina}}. The biological and environmental constraints might induce a new growth processes or cease a growth mechanism already existing in the system. Therefore, the inclusion or withdrawal of a growth mechanism from a system may be considered as deviation from the normal growth process due to some constraints. The allometry based biological growth is represented by the phenomenological class $U2$ {\cite {Castorina}}. Normal growth in a biological system shows a monotonous growth process by nature and attains a saturation level in course of time. A deviation from normal growth in terms of oscillation may be treated as adaptation or withdrawal of a new growth mechanism represented by different phenomenological terms other than $a+ba^2$ {\cite{Castorina}}. The growth of a tumour may be treated as a deviation from the normal growth due to some biological and perhaps environmental constraints. The coexistence of tumour growth along with normal growth of tissues within the same biological system and the beginning of tumour growth from normal growth may be treated as an indication of such deviation. The growth of a tumour shows such oscillatory nature in terms of its volume with relatively large amplitude {\cite {Deisboeck,Chignola}}.This communication is motivated by these phenomenological observation and aimed to describe phenomenologically oscillatory behavior of a system. \\
In the present communication, phenomenological approach based on the formalism presented by Castorina $et.$ $al.$ {\cite {Castorina}} is considered to describe the behaviour of a system showing oscillation.  Here, we propose two different phenomenological classes. One of them shows oscillation with constant amplitude (class-I). The other shows growth or decay in the amplitude of oscillation (class-II). The deviation from the proposed phenomenological class-I in terms of addition of a phenomenological term with the specified condition, might lead to a damped oscillatory growth (class-II). The emphasis is given to find out the role of coefficients involved in the phenomenological description of a dynamical system. The nature of variation of specific growth rate with different parameters is also studied. Absence of each phenomenological terms in the proposed phenomenological classes is also examined. Even a linear growth feature of the system may also be observed in absence of a phenomenological terms related to normal growth processes. In each case, the role of the phenomenological coefficients are examined in detail. The phenomenological term responsible for damping in the proposed class is also identified. This analysis is focused to help the experimentalists to decide whether their dataset, showing oscillatory nature, might be considerably expected to fit with these phenomenological classes.\\
The paper is organized as follows: In Sec. II, we would propose two phenomenological classes presenting oscillatory nature. In this connection, the classification scheme of phenomenological universalities proposed by Castorinal $et$ $al.$ {\cite {Castorina}} is discussed in brief. Different aspects of proposed phenomenological classes would be considered in Sec. III. Finally, we would conclude with our results in Sec IV.
\section{Phenomenological approach to oscillatory growth}
Following the formalism of Castorina $et$ $al.$ {\cite {Castorina}}, a string of data $y_j(t_j)$, showing a particular phenomenon that can be described by an ordinary differential equation, is represented in phenomenological approach in a following manner{\cite{Castorina}},
\begin{equation}
\frac{dY(t)}{dt}=\alpha(t)Y(t)
\label{e3:1}
\end{equation}
where, $\alpha(t)$ stands for specific growth rate of the variable $Y(t)$. Equation (\ref{e3:1}) can also be expressed in terms of adimensional variable as,
\begin{equation}
\frac{dy(\tau)}{d\tau}=a(\tau)y(\tau)
\label{e3:2}
\end{equation}
Where, $\tau=\alpha(0)t$, $y(\tau)=Y(t)/Y(0)$ and $a(\tau)=\alpha(t)/\alpha(0)$ are adimensional time, adimensional variable and adimensional specific growth rate respectively. It is assumed that $a(0)=y(0)=1$. An adimensional variable ($z$) is introduced in this connection so that,
\begin{equation}
z=\ln y
\label{e3:3}
\end{equation}
The time variation of $z$ is described by the following relations,
\begin{equation}
a=\frac{dz}{d\tau}
\label{e3:4}
\end{equation}
and,
\begin{equation}
\varphi(a)=-\frac{d^2z}{d\tau^2}
\label{e3:5}
\end{equation}
Again, $\varphi (a)$ can be expressed in terms of power series as,
\begin{equation}
\varphi(a)={\sum_0^\infty} b_n a^n
\label{e3:6}
\end{equation}
From equation (\ref{e3:2}), (\ref{e3:3}) and (\ref{e3:5}), the following expression can be derived,
\begin{equation}
z=\int ad\tau+ constant
\label{e3:7}
\end{equation}
Different forms of $\varphi$ would generate different types of growth equation. Each term in the right hand side of the equation (\ref{e3:6}) represents different types of growth mechanism involved in the system {\cite {Castorina}}. These mechanisms might be independent with respect to each other or they might be mutually dependent. In case of Gompertz-type growth, $b_0=0$ and $b_n=0$ for $n\geq2$. It indicates that the growth mechanism corresponding to the term $b_1a$ is not dependent on the other growth mechanisms found in nature. In case of West-type allometry based biological growth process, $b_0=0$ and $b_n=0$ for $n\geq3$. The growth mechanism corresponding to the terms $b_1a$ and $b_2a^2$ are simultaneously found in this type of system. Thus, the sole existence of the growth mechanism corresponding to $b_1a$ is reported and co-existence of growth mechanisms corresponding to the terms $b_1a$ and $b_2a^2$ are also found in nature. The sole existence of the growth mechanism corresponding to the term $b_2a^2$ or a coexistence with the terms other than $b_1a$ is still not reported or not even considered from the theoretical point of view in phenomenological approach. Here, such a possibility is explored from phenomenological point of view.\\
In the phenomenological class represented by $\varphi=b_0+b_2a^2$ with $b_1=0$ and all $b_n=0$ for $n>2$, the analytical solution of $y$ is expressed as,
\begin{equation}
y=A^{\frac{1}{\alpha \beta}}\mid cos[\alpha \tau - \theta]\mid ^{\frac{1}{\alpha \beta}}
\label{e3:8}
\end{equation}
Where, $A=\frac{1}{\mid cos \theta\mid }$, $\theta = \arctan \beta$, $\alpha = \sqrt{b_0b_2}$, $\beta =\sqrt{\frac{b_2}{b_0}}$. Here, $\alpha$ may be treated as frequency of oscillation.\\
The solution can also be achieved in terms of $sine$ function as,
\begin{equation}
y=A^{\frac{1}{\alpha \beta}}\mid sin[\alpha \tau + \xi]\mid ^{\frac{1}{\alpha \beta}}
\label{e3:9}
\end{equation}
Where, $A=\frac{1}{\mid sin \xi\mid }$, $\xi = \cot^{-1} \beta$.\\
Though the solution can be expressed in terms of $sine$ function as well as $cosine$ function, while the $cosine$ function would be considered to describe the behaviour of the system in this communication. It is verified that the same features of the system can be extracted in both cases.\\
The specific growth rate of this phenomenological class is expressed as,
\begin{equation}
a=\frac{1}{\beta}\tan (\theta-z)
\label{e3:10}
\end{equation}
Where, $z=\sqrt{b_0b_2}\tau$.\\
It is not possible to describe the behaviour of the system with the help of a single differential equation. It is because of the fact that $y$ is not differentiable when $[\alpha \tau - \theta]=\frac{\pi}{2}, \frac{3\pi}{2}, \frac{5\pi}{2}, ........$. Therefore, the formation of segmented or piece-wise differential equations are possible that represent the behaviour of the system. The differential equation given below,
\begin{equation}
\frac{1}{y}\frac{d^2y}{d\tau^2}+\frac{b_2-1}{y^2}(\frac{dy}{d\tau})^2=b_0
\label{e3:11}
\end{equation}
is valid for the interval $(2n+1)\frac{\pi}{2}<[\alpha \tau - \theta]<(2n+3)\frac{\pi}{2}$, where $n=0, 2, 4, ......$ ~. For the interval $0<[\alpha \tau - \theta]<\frac{\pi}{2}$ and $(2m+1)\frac{\pi}{2}<[\alpha \tau - \theta]<(2m+3)\frac{\pi}{2}$ (where, $m=1, 3, 5, .....$), the differential equation is given by,
\begin{equation}
\frac{1}{y}\frac{d^2y}{d\tau^2}+\frac{b_2-1}{y^2}(\frac{dy}{d\tau})^2=-b_0
\label{e3:12}
\end{equation}
For the phenomenological class corresponding to $\varphi=b_0+b_1a+b_2a^2$ with all $b_n=0$ for $n>2$ and satisfying the condition $b_1^2<4b_0b_2$; i.e; $zeros$ of the function $\varphi$ is complex; the system is governed by the following expression,
\begin{equation}
y=Pexp(-\sigma t)\mid \cos(\omega\tau - \gamma)\mid ^{\frac{1}{b_2}}
\label{e3:13}
\end{equation}
Where, $P=[\frac{b_1^2+4b_2(b_2+b_1+b_0)}{4b_0b_2-b_1^2}]^{\frac{1}{2b_2}}$, $\sigma=\frac{b_1}{2b_2}$, $\omega=\sqrt{b_0b_2-\frac{b_1^2}{4}}$, $\tan \gamma=\frac{b_1+2b_2}{\sqrt{4b_0b_2-b_1^2}}$.\\
In this case, $\sigma = \frac{b_1}{2b_2}$, may be termed as damping factor of the oscillating system, and $\omega =\sqrt{b_0b_2-\frac{b_1^2}{4}}$, may be termed as angular frequency of the oscillation,
\section{Discussions}
\begin{figure}
  \centering
  \includegraphics[width=3in, height=3in]{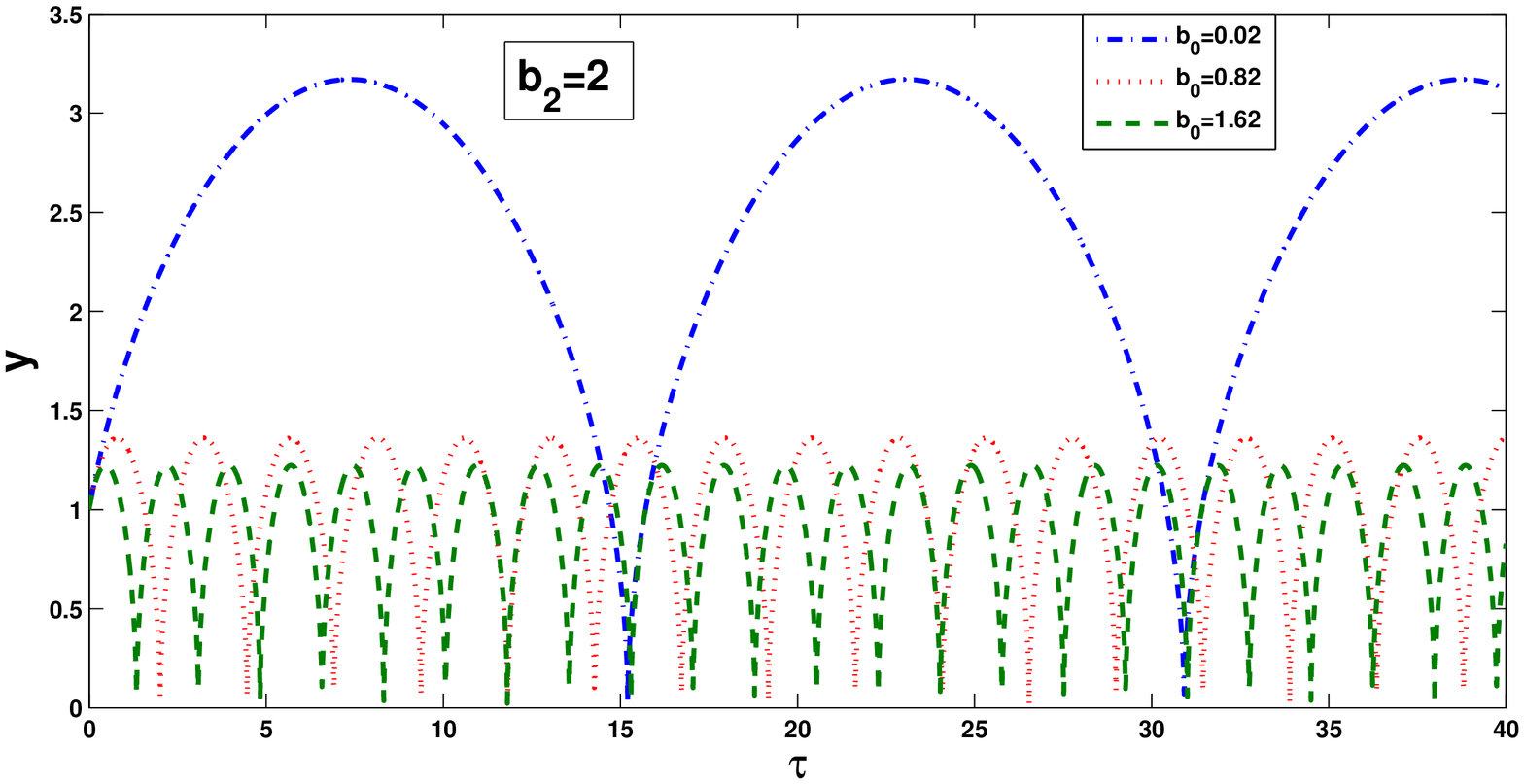}\\
  \caption{(Colour online) Plots of oscillatory nature of the variable $y$ with $b_1=0$ and $b_2=2$. From top to the bottom the values of the parameter $b_0$ are $0.02$, $0.82$  and $1.62$.}
\label{fig3:1}
\end{figure}

\begin{figure}
  \centering
  \includegraphics[width=3in, height=3in]{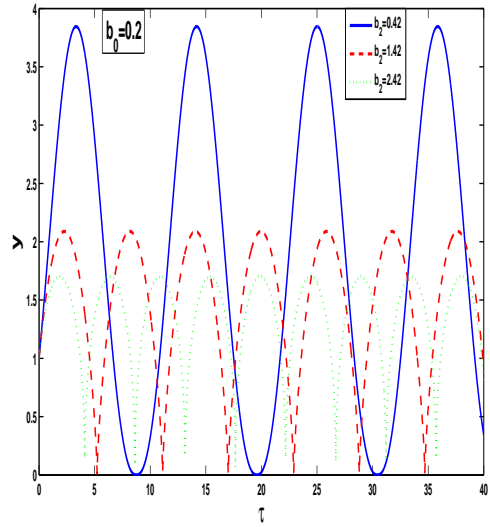}\\
  \caption{(Colour online) Plots of oscillatory nature of the variable $y$ with $b_1=0$ and $b_0=0.2$. From top to the bottom the values of the parameter $b_2$ are $0.42$, $1.42$  and $2.42$.}
\label{fig3:2}
\end{figure}

\begin{figure}
  \centering
  \includegraphics[width=3in, height=3in]{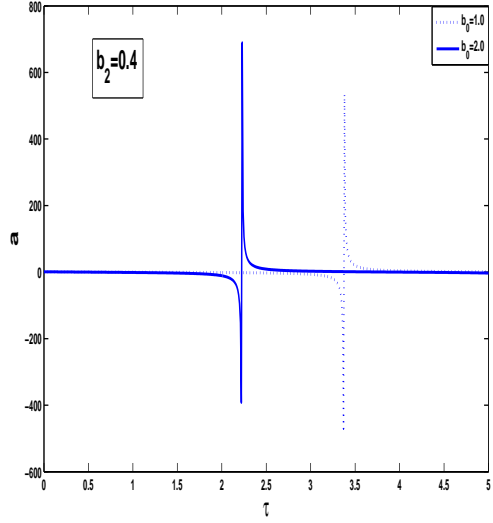}\\
  \caption{(Colour online) Curves of specific growth rate with $b_1=0$ and $b_2=0.4$. From left to right the values of the parameter $b_0$ are $2.0$ and $1.0$. os11}
\label{fig3:3}
\end{figure}

\begin{figure}
  \centering
  \includegraphics[width=3in, height=3in]{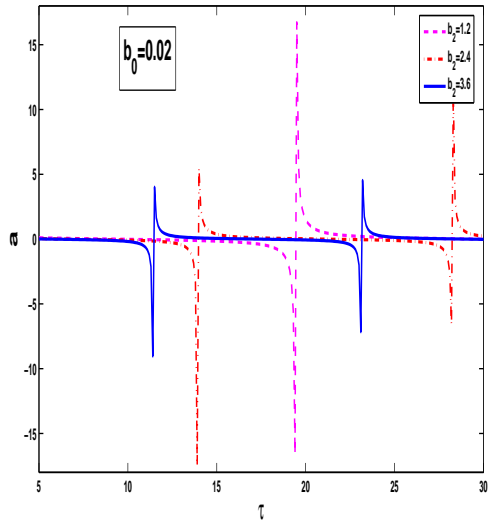}\\
  \caption{(Colour online) Curves of specific growth rate with $b_0=0.02$ and $b_1=0$. From left to right the values of the parameter $b_2$ are $3.6$, $2.4$ and $1.2$.}
\label{fig3:4}
\end{figure}

\begin{figure}
  \centering
  \includegraphics[width=3in, height=3in]{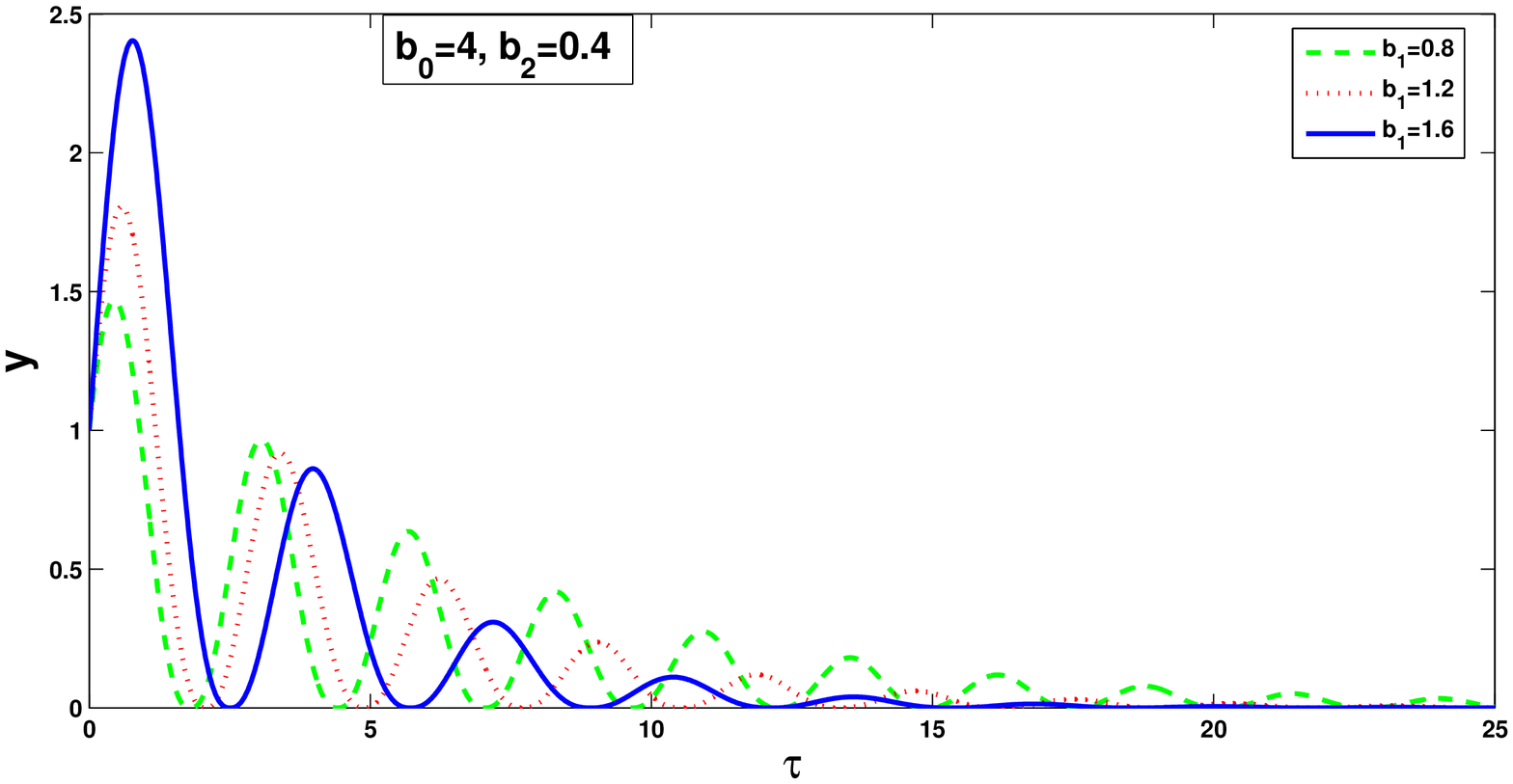}\\
  \caption{(Colour online) Plots of damped-oscillatory nature of the variable $y$ with $b_0=4.0$ and $b_2=0.4$. From top to the bottom the values of the parameter $b_1$ are $0.8$, $1.2$ and $1.6$.}
\label{fig3:5}
\end{figure}

\begin{figure}
  \centering
  \includegraphics[width=3in, height=3in]{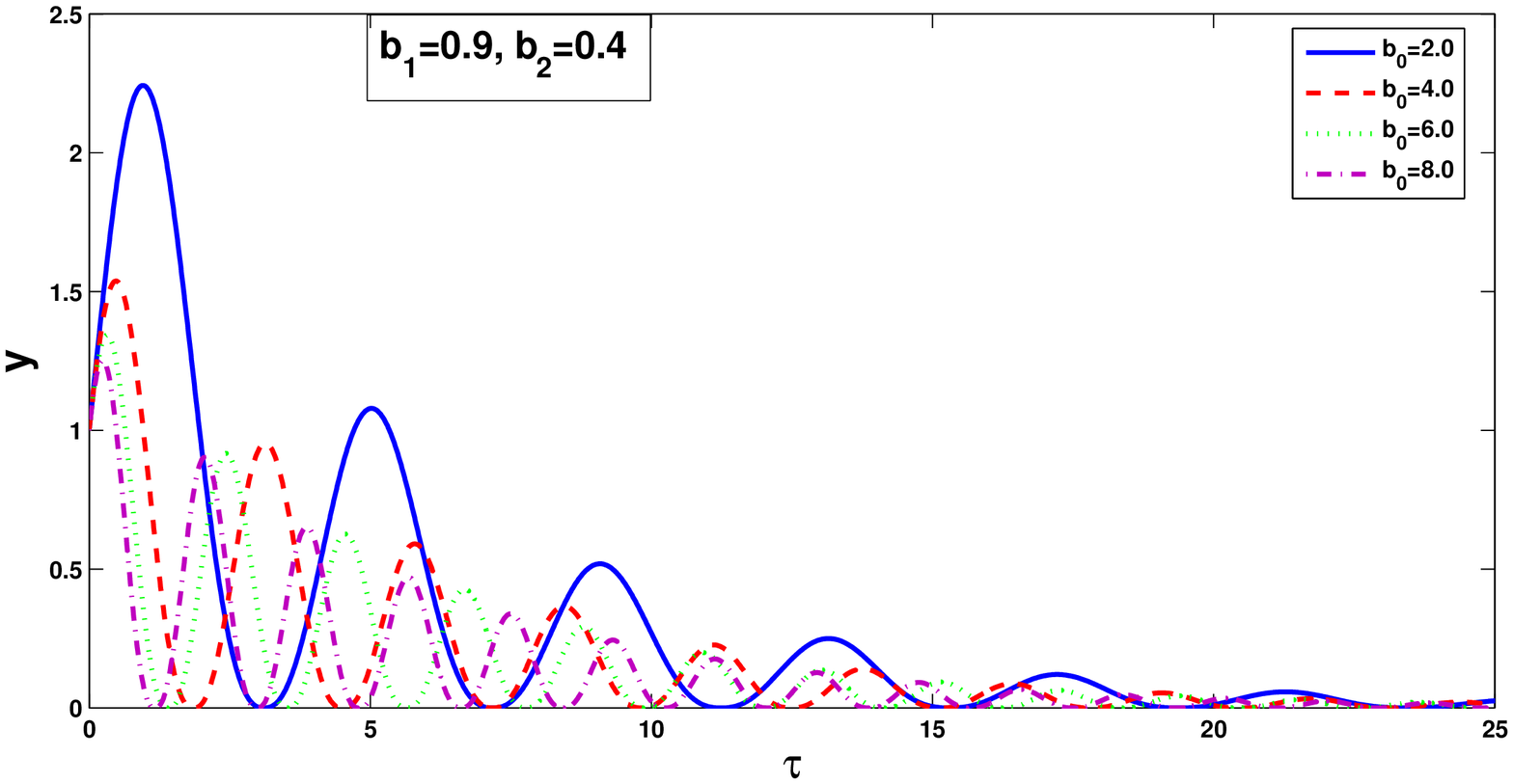}\\
  \caption{(Colour online) Plots of damped-oscillatory nature of the variable $y$ with $b_1=0.9$ and $b_2=0.4$. From top to the bottom the values of the parameter $b_0$ are $2.0$, $4.0$, $6.0$ and $8.0$.}
\label{fig3:6}
\end{figure}

\begin{figure}
  \centering
  \includegraphics[width=3in, height=3in]{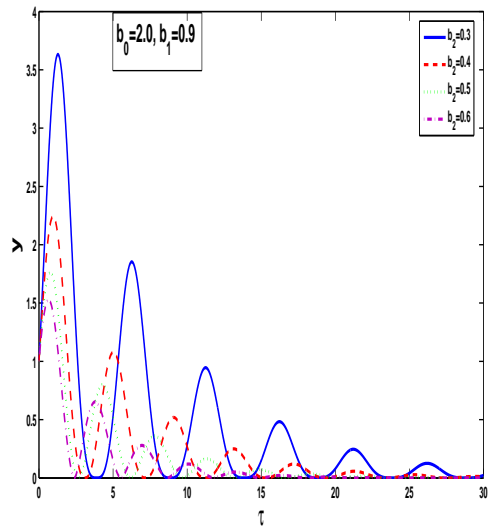}\\
  \caption{(Colour online) Plots of damped-oscillatory nature of the variable $y$ with $b_0=2.0$ and $b_1=0.9$. From top to the bottom the values of the parameter $b_2$ are $0.3$, $0.4$, $0.5$ and $0.6$.}
\label{fig3:7}
\end{figure}

\begin{figure}
  \centering
  \includegraphics[width=3in, height=3in]{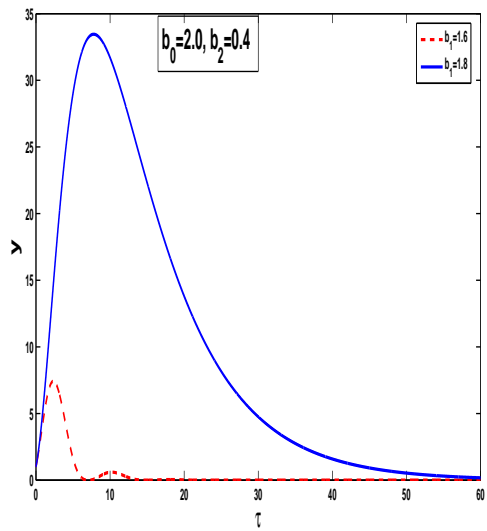}\\
  \caption{(Colour online) Plots of over-damped nature of the variable $y$ with $b_0=2.0$ and $b_2=0.4$. From top to the bottom the values of the parameter $b_1$ are $1.6$ and $1.8.$. The condition defined by $b_0=2.0$, $b_1=1.8$ and $b_2=0.4$ is just above the threshold of the condition $b_1^2<4b_0b_2$ and represents a behaviour similar to overdamped nature of a classical oscillator.}
\label{fig3:8}
\end{figure}

In the proposed class of phenomenological universalities defined by $\varphi=b_0+b_2a^2$;~ $\alpha$, being a function of $b_0$ and $b_2$, and $A^{\frac{1}{\alpha \beta}}$ can be treated as the angular frequency and amplitude of oscillation respectively. It is found that the amplitude of oscillation of the system is affected by the variation of both $b_0$ and $b_2$. In both cases, amplitude decreases with the increase of $b_0$ or $b_2$. In case of change in frequency of oscillation, $b_0$ and $b_2$ play same role, but the change in frequency is more affected by the change in $b_0$ than the change in $b_2$. The frequency of oscillation increases with the increase in $b_0$ and $b_2$. The change in nature of oscillation with the change in $b_0$ and $b_2$ is represented in figure (\ref{fig3:1}) and (\ref{fig3:2}). The nature of oscillation can also be characterized by the sharpness of oscillation that does not depend on $b_0$. It is totally controlled by $b_2$ (as it is observed in graphical representation). The sharpness of the characteristic curve increases with the decrease in $b_2$. The change in sharpness of oscillation with the change in $b_2$ is represented in figure (\ref{fig3:2}). No such behaviour is observed in case of $b_0$.This proposed phenomenological class might be useful to describe phenomenologically sustained oscillations observed in different physical systems {\cite{dane,pullela,tu2005,tu2007}}.\\
The change in specific growth rate with the change in $b_0$ and $b_2$ are represented in figure (\ref{fig3:3}) and figure (\ref{fig3:4}). It is found that the frequency of oscillation of specific growth rate increases with the increase in $b_0$ and $b_2$. As a consequence, the time period corresponding to the change in specific growth rate decreases with the increase in the values of phenomenological coefficients. The frequency of oscillation of $y$ is identical with the frequency of oscillation of specific growth rate, as expected from equation (\ref{e3:8}) and (\ref{e3:10}).\\
The phenomenological class corresponding to $\varphi=b_0+b_1a+b_2a^2$ with the condition $b_1^2<4b_0b_2$ leads to damped oscillatory nature of a growing system when $b_1$ and $b_2$ are greater than zero.. The amplitude of oscillation is given by $P exp(-\sigma \tau)$. Therefore, the exponential decay in amplitude is governed by the magnitude of $b_1$ and $b_2$, as it is expected from equation (\ref{e3:13}). The decay in amplitude increases with increase in $b_1$ and with the decrease in $b_2$, as shown in the figure (\ref{fig3:5}) and figure (\ref{fig3:7}). The decay does not depend on $b_0$ as it is not function of $b_0$. Thus, it can be concluded that the damping of the system depends on $b_1$ and $b_2$. It is practically governed by $b_1$ because of the fact that an undamped system may be considered for either $b_1=0$ or $b_2\rightarrow \infty$ but that is not feasible in reality.  The magnitude of $P$ increases with the increase in $b_1$. But, the same decreases with the increase in $b_0$ and $b_2$. The frequency of the damped oscillatory system represented by this phenomenological class may be represented by $\omega$, as it is found in equation (\ref{e3:13}). It increases with the increase in $b_0$ and $b_2$ (shown in figure (\ref{fig3:6}) and figure (\ref{fig3:7})) but the same decreases with the increase in $b_1$ (shown in figure (\ref{fig3:5})).It is interesting to note that the damped frequency ($\omega$) of oscillation is less than the undamped frequency ($\alpha$) of oscillation. The difference between $\alpha$ and $\omega$ depends on $b_1$. The amplitude of oscillation shows exponential growth for $b_1<0$ and $b_2>0$. This proposed phenomenological class might be helpful to describe oscillatory behaviour of a system accompanied by damping (may be positive or negative).\\
The characteristic equation of undamped oscillation corresponding to the phenomenological class $\varphi=b_0+b_2a^2$ could be obtained from the characteristic solution of the phenomenological class represented by $\varphi=b_0+b_1a+b_2a^2$ by considering $b_1=0$. In a similar fashion, the undamped frequency ($\alpha$) is obtained by considering $b_1=0$ in the expression of damped frequency ($\omega$) which is less than the undamped frequency. Therefore, the change in frequency due to damping is solely determined by the term $b_1$, that also controls the decay in amplitude due to damping. Thus, it can be concluded that the mechanism involved in the system represented by the term $b_1a$ is independent with respect to other mechanisms and solely responsible for damping. In absence of the terms $b_0$ and $b_1$, the phenomenological class is represented by $\varphi=b_2a^2$ and corresponding growth function is given by,
\begin{equation}
y=(b_2\tau +1)^{\frac{1}{b_2}}
\label{e3:14}
\end{equation}
The same can also be derived from equation (\ref{e3:13}). In this case, $b_2=1$ leads to a linear growth of the system. A system showing always linear growth with respect to time is not normally observed in nature.\\
When the values of $b_0$, $b_1$ and $b_2$ are at the threshold condition $b_1^2<4b_0b_2$, the oscillatory nature along with damping is about to cease, as represented in the figure (\ref{fig3:8}). It is also found that the behaviour of the system is similar to the critically damped oscillatory nature of a classical oscillator when the the values of the phenomenological coefficients are just above the threshold condition $b_1^2<4b_0b_2$. This nature is represented in the figure (\ref{fig3:8}).
The behaviours of the damped oscillation and undamped oscillation of the phenomenological classes are very similar to the behaviours observed in case of a classical oscillations. In case of a classical oscillator, the undamped oscillation is characterised by the mass of the system, the stiffness factor or force constant and the damping determined by the interaction of the system with the environment. The withdrawal of that interaction causing damping generates an undamped oscillation which is also found in this phenomenological approach. The damping of a classical oscillator depends on the property of the surrounding medium. Mass of a classical oscillator contributes in determining the damping and frequency of oscillation. Therefore, as an analogy, it can be concluded that the term $b_1$ plays the same role as the interaction of the system with the environment in case of classical oscillator. The terms, $b_0$ and $b_2$, determine undamped frequency of oscillation that is governed by force constant and mass of a classical oscillator.
\section{Conclusions}
In the present communication, the oscillating behaviour of a growing system is considered in the framework of phenomenological approach. The variation in oscillating behaviour is analyzed with the variation of different phenomenological coefficients and is represented graphically. The variation in specific growth rate is also studied. The phenomenological class representing damped oscillatory nature is identified with the conditional relationship between the phenomenological coefficients. In this connection, an analogy between the behaviour of damped classical oscillation and the damped oscillation in phenomenological approach is considered. The term solely responsible for damping is also pointed out. An effort is given to identify the physical significance of each term involved in the phenomenological description from this analogy. In a nutshell, the proposed phenomenological description is able to address the oscillation of a system with constant amplitude. The phenomenological classes may be useful to describe the behaviour of oscillatory physical systems {\cite{pullela,dane,Deisboeck,Chignola}}. It may be beneficial to address oscillatory growth of a tumor. These proposed phenomenological classes might be helpful for the experimentalists to consider their dataset from a totally different point of view and to extract a suitable model for describing the temporal evolution of the system.


\end{document}